\documentclass[a4paper,twocolumn]{article}
\usepackage{fullpage}
\usepackage{amsfonts}
\usepackage{gastex}

\newtheorem{thm}{Theorem}

\begin{document}

\title{The reachability problem for vector addition systems \\
       with a stack is not elementary}
\author{Ranko Lazi\'c}
\date{DIMAP, Department of Computer Science, University of Warwick, UK}

\maketitle

\begin{abstract}
By adapting the iterative yardstick construction of Stockmeyer,
we show that the reachability problem for vector addition systems
with a stack does not have elementary complexity.
As a corollary, the same lower bound holds for the satisfiability problem
for a two-variable first-order logic on trees in which 
unbounded data may label only leaf nodes.
Whether the two problems are decidable remains an open question.
\end{abstract}

\section{Introduction}

\begin{figure*}
\setlength{\unitlength}{.7em}
\begin{center}
\begin{picture}(60,27)(0,-1)
\node[Nadjust=wh](VAS)(30,2.5)
  {\begin{tabular}{c}
   VAS \\ \hline \hline
   decidable \\ \hline
   \textsc{ExpSpace}-hard
   \end{tabular}}
\node[Nadjust=wh](BVAS)(22.5,12.5)
  {\begin{tabular}{c}
   BVAS \\ \hline \hline
   ? \\ \hline
   2\textsc{ExpSpace}-hard
   \end{tabular}}
\node[Nadjust=wh](PVAS)(37.5,12.5)
  {\begin{tabular}{c}
   PVAS \\ \hline \hline
   decidable \\ \hline
   \textsc{ExpSpace}-hard
   \end{tabular}}
\node[Nadjust=wh](SVAS)(37.5,22.5)
  {\begin{tabular}{c}
   SVAS \\ \hline \hline
   ? \\ \hline
   \textbf{not elementary}
   \end{tabular}}
\node[Nadjust=wh](VAS2)(12.5,2.5)
  {\begin{tabular}{c}
   FO$^2(+1, <, \sim)$ \\
   data words
   \end{tabular}}
\node[Nadjust=wh](BVAS2)(5,12.5)
  {\begin{tabular}{c}
   FO$^2(+1, <, \sim)$ \\
   data trees
   \end{tabular}}
\node[Nadjust=wh](PVAS2)(55,12.5)
  {\begin{tabular}{c}
   FO$^2(+1, <, \prec, \sim)$ \\
   bounded-depth \\
   data trees
   \end{tabular}}
\node[Nadjust=wh](SVAS2)(55,22.5)
  {\begin{tabular}{c}
   FO$^2(+1, \prec, \sim)$ \\
   leaf-data trees
   \end{tabular}}
\drawedge[AHnb=0](VAS,BVAS){}
\drawedge[AHnb=0](VAS,PVAS){}
\drawedge[AHnb=0](PVAS,SVAS){}
\drawedge[curvedepth=1.5](VAS,VAS2){}
\drawedge[curvedepth=1.5](VAS2,VAS){}
\drawedge(BVAS,BVAS2){}
\drawedge(PVAS2,PVAS){}
\setlength{\unitlength}{1mm}
\drawedge[linewidth=0.28](SVAS,SVAS2){}
\end{picture}
\end{center}
\caption{Reachability for extensions of VAS and 
         satisfiability for 2-variable FO with data}
\label{f:1}
\end{figure*}
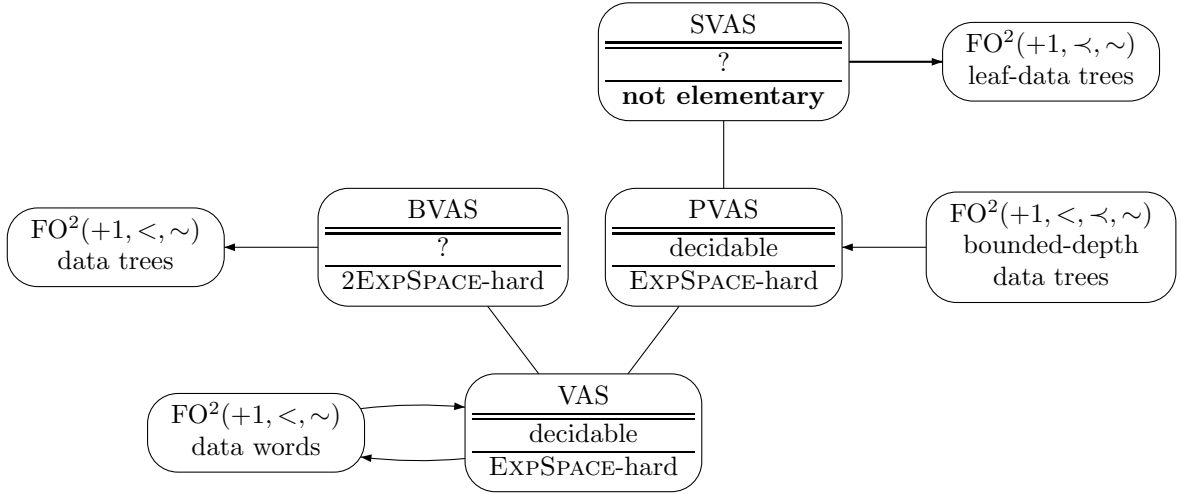

Before presenting details of this small contribution to
the on-going investigation of complexity-theoretic properties
of vector addition systems, their extensions and related logics
on words and trees with unbounded data, we provide a brief overview
of most-closely related research.
A diagrammatic summary is in Figure~\ref{f:1},
where boldface and a thicker line indicate the new results.

\paragraph*{VAS.}

Vector addition systems, or equivalently Petri nets,
are a fundamental and established model of concurrency.
They support an active and broad research community,
with long-standing links to industry, 
where VAS are an important modelling formalism 
and associated tools are extensively used.  

In spite of several decades of research, 
the computational complexity of the reachability problem for VAS 
remains one of the most well-known open questions 
in theoretical computer science.
While decidability was still unknown, Lipton made initial progress
on the problem by showing it \textsc{ExpSpace}-hard \cite{Lipton76},
which is still the highest known lower bound.
A few years later, Mayr showed the problem decidable \cite{Mayr84}.
Although his proof was subsequently substantially simplified by
Kosaraju \cite{Kosaraju82}, Lambert \cite{Lambert92} and
in a remarkable recent series of articles by Leroux \cite{Leroux12},
it is still unknown even whether there exists 
a primitive recursive algorithm for the problem.

\paragraph*{Branching VAS.}

Whereas computations of VAS are words of vectors of
natural numbers, BVAS are a natural generalisation
whose computations are trees of such vectors.
Although their reachability problem has been shown 
inter-reducible with the emptiness problem for
multiple-valued linear index grammars \cite{Rambow94,Schmitz10},
and with the provability problem for 
multiplicative exponential linear logic \cite{deGroote&Guillaume&Salvati04},
the decidability status remains an open question.
However, curiously, a lower bound that is two notches above
adding alternation to Lipton's result, namely 2\textsc{ExpSpace}-hardness, 
was recently shown~\cite{Lazic10}.

\paragraph*{Priority VAS.}

Equipping two counters (in Petri-speak, places) with zero tests,
of course, makes VAS as powerful as Minsky machines and
the reachability problem undecidable.
It has turned out, though, that the Mayr-Kosaraju-Lambert proof
can be extended when only one counter may be tested for zero.
In fact, Reinhardt has obtained a highly non-trivial proof of
an even more general result: that reachability is decidable for
\emph{PVAS}, where one may test whether 
all counters from any one of a series of sets 
$C_1 \subseteq C_2 \subseteq \cdots C_k$ are zero
\cite{Reinhardt08}.
So far with one zero-testable counter,
Bonnet has succeeded in greatly simplifying Reinhardt's proof
along the lines of Leroux \cite{Bonnet11}.

Let us say that PVAS whose series of zero-testable sets of counters
have length $k$ are of \emph{index}~$k$.

\paragraph*{Stack VAS.}

Another natural extension of VAS is 
to allow them to use a stack over a finite alphabet.
Equivalently to these systems, which we call SVAS and whose motivations 
include modelling software with integer variables and call-return procedures,
one may consider intersections of VAS languages and context-free languages.
For an SVAS in that alternative presentation, 
let us say that it is of \emph{index} $k$ if and only if
the context-free language is of index $k$, 
i.e.\ there is a context-free grammar such that 
every word in the language has a derivation whose 
every step contains at most $k$ non-terminal symbols.
Atig and Ganty have recently shown that finite-index SVAS
are essentially equivalent to PVAS:
every index-$k$ SVAS can be simulated by an index-$k$ PVAS,
and every index-$k$ PVAS can be simulated by an index-$(k + 1)$ SVAS
\cite{AtigGanty11}.
Incidentally, that seems to be the only interesting known relationship
among BVAS, PVAS and SVAS.

The reachability problem for finite-index SVAS is consequently decidable
since it is decidable for PVAS.  
Although decidability for unrestricted SVAS remains an open question,
we make some progress here in the opposite direction,
obtaining that the problem is not elementary.
That puts SVAS in contrast to BVAS, for which decidability is also unknown 
but so far there is only an elementary lower bound \cite{Lazic10}.

\paragraph*{Coverability.}

The well-known coverability problem for VAS and their extensions 
corresponds to ``control-state reachability'': 
it asks whether a given system can reach a configuration that is pointwise 
(i.e., for each counter) greater than or equal to a given configuration.
Lipton's and Rackoff's classical results show that coverability for VAS
is \textsc{ExpSpace}-complete \cite{Lipton76,Rackoff78},
and by building on those works, Demri et al.\ have shown 
2\textsc{ExpTime}-completeness of the problem for BVAS \cite{Demrietal12}.

Unfortunately, for PVAS and SVAS, there is no hope for such results,
since for both classes of systems, there are straightforward reductions 
of reachability to coverability.

\paragraph*{2-variable FO on data words and data trees.}

Partly motivated by verification of concurrent systems 
and by querying of XML databases, in recent years there has been
extensive research in logics on data words and data trees.
In addition to letters from a finite alphabet as classically,
the latter structures have labels from an infinite domain,
which are called \emph{data} and on which only certain operations are available.
In fact, typically, the data can only be compared for equality, 
and that is the only operation we consider here.

Remarkably, there are several connections between, on one hand, 
VAS and their extensions that we have introduced, and on the other hand, 
two-variable first-order logics on data words and data trees.
For positions $x$ and $y$ of a data word, 
the logics have navigational predicates $y = x + 1$ and $x < y$, 
as well as equality of data labels $x \sim y$.
On data trees, where variables range over nodes,
navigational predicates are either vertical (``child'' and ``descendant''),
or horizontal (``next sibling'' and ``following sibling''),
or compare nodes for positions in the pre-traversal (``document order'').

On data words, Boja\'nczyk et al.\ \cite{Bojanczyketal11}
showed that the satisfiability problem for such a logic
reduces in doubly-exponential time to the reachability problem for VAS,
and is therefore decidable.  Moreover, they exhibited
a polynomial-time converse reduction, and so Lipton's lower bound
carries over to the logic.

On data trees, the picture is more complicated.
Already without document order, Boja\'nczyk et al.\ \cite{Bojanczyketal09}
observed that the satisfiability problem is at least as hard as 
the reachability problem for BVAS (whose decidability is open),
but obtained decidability by disallowing also the transitive
navigational predicates (``descendant'' and ``following sibling'').
Another way of getting decidability was found by 
Bj\"orklund and Boja\'nczyk \cite{Bjorklund&Bojanczyk07a}:
no restrictions on the navigational predicates are required 
provided the depth of data trees is bounded.
With that assumption, they showed how to reduce satisfiability to
the reachability problem for PVAS.

An alternative restriction on data trees suggests itself:
that data labels be allowed only on leaf nodes.
Although decidability of the full 2-variable FO on such structures
remains open, we show that even without the ``descendant'' and
``following sibling'' predicates, satisfiability is at least as hard as
the reachability problem for SVAS, and so is not elementary.

\section{Lower bound}

It is convenient for our purposes to formalise SVAS as
programs which operate on non-negative counters and a finite-alphabet stack.
More precisely, we define them as 
finite sequences of commands which may be labelled,
where a command is one of:
an increment of a counter ($x := x + 1$),
a decrement of a counter ($x := x - 1$),
a push ($\mathtt{push}\ a$),
a pop ($\mathtt{pop}\ a$),
a non-deterministic jump to one of two labelled commands 
($\mathtt{goto}\ L\ \mathtt{or}\ L'$),
or termination ($\mathtt{halt}$).
Initially, all counters have value $0$ and the stack is empty.
Whenever a decrement of a counter with value $0$ or an erroneous pop 
is attempted, the program aborts.
In every program, $\mathtt{halt}$ occurs only as the last command.

The reachability problem can now be stated as follows: given an SVAS, 
does it have a computation which reaches the $\mathtt{halt}$ command 
with all counters being $0$ and the stack being empty?

\begin{thm}
The reachability problem for SVAS is not elementary.
\end{thm}

The proof is by reducing from 
the $(2 \Uparrow n)$-bounded halting problem for 
\emph{counter programs} with $n$ commands,
where:
\begin{itemize}
\item
for $k \in \mathbb{N}$,
the \emph{tetration} operation $b \Uparrow k$ is defined by
$b \Uparrow 0 = 1$ and $b \Uparrow (k + 1) = b^{b \Uparrow k}$;
\item
the counter programs are defined like SVAS, 
except that they have no stack, 
have only deterministic jumps ($\mathtt{goto}\ L$),
but can test counters for zero 
($\mathtt{if}\ x = 0\ \mathtt{then}\ L\ \mathtt{else}\ L'$);
\item
the $(2 \Uparrow n)$-bounded halting problem asks whether
$\mathtt{halt}$ is reachable by a computation during which 
all counter values are at most $2 \Uparrow n$.
\end{itemize}

Given such a counter program $\mathcal{C}$,
we construct in time polynomial in $n$
an SVAS $\mathcal{S}(\mathcal{C})$ which simulates $\mathcal{C}$
as long as its counters do not exceed $2 \Uparrow n$.
As in Stockmeyer's yardstick construction \cite{Stockmeyer74},
the idea is to bootstrap the ability to simulate zero tests
of counters that are bounded by $2 \Uparrow 1$,
$2 \Uparrow 2$, \ldots, $2 \Uparrow n$.

More precisely, for each counter $x$ of $\mathcal{C}$,
$\mathcal{S}(\mathcal{C})$ has a pair of counters $x$ and $\bar{x}$,
on which it maintains the invariant $x + \bar{x} = 2 \Uparrow n$.
Thus, every increment of $x$ in $\mathcal{C}$ is translated to 
$x := x + 1;\ \bar{x} := \bar{x} - 1$ in $\mathcal{S}(\mathcal{C})$,
and similarly for decrements.

For every zero test of $x$ in $\mathcal{C}$,
$\mathcal{S}(\mathcal{C})$ uses auxiliary counters $s_n$ and $\bar{s}_n$,
for which it also maintains $s_n + \bar{s}_n = 2 \Uparrow n$.
Moreover, we assume that $s_n = 0$ 
at the start of each zero-test simulation.
The simulation begins by $\mathcal{S}(\mathcal{C})$ 
transferring some part of $\bar{x}$ to $s_n$
(while preserving the invariants).
It then calls a procedure $\mathit{Dec}_n$
which decrements $s_n$ exactly $2 \Uparrow n$ times.
For the latter to be possible, $x$ must have been $0$.
Otherwise, or in case not all of $\bar{x}$ was transferred to $s_n$,
the procedure can only abort.
When $\mathit{Dec}_n$ succeeds,
the initial values of $x$ and $\bar{x}$ are reversed,
so to finish the simulation, 
everything is repeated with $x$ and $\bar{x}$ swapped.

The main part of the construction is 
implementing $\mathit{Dec}_k$ for $k = 1, 2, \ldots, n$.
Assuming that $\mathit{Dec}_k$ which decrements $s_k$ 
exactly $2 \Uparrow k$ times and maintains $s_k + \bar{s}_k = 2 \Uparrow k$
has been implemented for some $k < n$,
$\mathit{Dec}_{k + 1}$ consists of performing the following
by means of $s_k$, $\bar{s}_k$ and $\mathit{Dec}_k$:
\begin{itemize}
\item
push exactly $2 \Uparrow k$ zeros onto the stack;
\item
keep incrementing the $(2 \Uparrow k)$-digit binary number
that is on top of the stack until no longer possible, 
and decrement $s_{k + 1}$ for each such increment;
\item
pop $2 \Uparrow k$ ones that are on top of the stack,
and decrement $s_{k + 1}$ once more.
\end{itemize}

By a similar pattern, starting with all counters having value $0$,
$\mathcal{S}(\mathcal{C})$ can initialise 
each auxiliary counter $\bar{s}_k$ to $2 \Uparrow k$,
and each $\bar{x}$ to $2 \Uparrow n$,
as required.

\section{Reduction to logic}

Let \emph{leaf-data forests} be data forests in which 
data labels are present only at leaf nodes.
More precisely, they are finite forests such that:
\begin{itemize}
\item
the root nodes are linearly ordered;
\item
each node is either a leaf, 
or its children and their descendants form a leaf-data forest;
\item
each node has a label from a finite alphabet $\Sigma$;
\item
each leaf node also has a label from an infinite domain (say, $\mathbb{N}$).
\end{itemize}

Now, let FO$^2(+1, \prec, \sim)$ denote 
the two-variable first-order logic on leaf-data forests
that has the following predicates:
\begin{itemize}
\item
a unary predicate for each letter from $\Sigma$;
\item
$x \downarrow y$ ($y$ is a child of $x$) and 
$x \rightarrow y$ ($y$ is the next sibling of $x$);
\item
$x \prec y$ 
($x$ and $y$ are leaves, and $x$ precedes $y$ in the document order);
\item
$x \sim y$ 
($x$ and $y$ are leaves with the same data label).
\end{itemize}

\begin{thm}
The reachability problem for SVAS
is reducible in logarithmic space to
the satisfiability problem for FO$^2(+1, \prec, \sim)$ on leaf-data forests.
\end{thm}

The proof is based on encoding SVAS computations as leaf-data forests.
In the latter, their tree structure is used 
to represent the evolution of the stack,
and data labels are employed for keeping track of counter values.

More concretely, suppose $\mathcal{S}$ is an SVAS.
We show how to compute in logarithmic space
a sentence $\phi(\mathcal{S})$ of FO$^2(+1, \prec, \sim)$
whose models are exactly leaf-data forests that encode
in the following manner computations of $\mathcal{S}$
that halt with all counters $0$ and the stack empty:
\begin{itemize}
\item
the computation that such a leaf-data forest encodes can be 
obtained by traversing the forest so that each internal node
is visited once before its children (generating a push) and
for a second time after its children (generating the corresponding pop);
\item
each leaf node is labelled either by a jump command,
or by an increment or a decrement, and in the latter cases,
mutually distinct data labels are used to distinguish among
increments of the same counter and to match them to its decrements;
\item
each internal node is labelled by a pair consisting of
a push command and the pop command that corresponds to it
in the computation.
\end{itemize}

\section*{Acknowledgements}

I am grateful to Miko{\l}aj Boja\'nczyk and Anca Muscholl
for outlining to me the reduction from SVAS reachability to
FO$^2(+1, \prec, \sim)$ satisfiability on leaf-data trees.

\bibliographystyle{abbrv}
{\footnotesize \bibliography{svas}}

\begin{thebibliography}{10}

\bibitem{AtigGanty11}
M.~F. Atig and P.~Ganty.
\newblock Approximating {P}etri net reachability along context-free traces.
\newblock In {\em FSTTCS}, volume~13 of {\em LIPIcs}, pages 152--163. Schloss
  Dagstuhl, 2011.

\bibitem{Bjorklund&Bojanczyk07a}
H.~Bj\"orklund and M.~Boja\'nczyk.
\newblock Bounded depth data trees.
\newblock In {\em ICALP}, volume 4596 of {\em LNCS}, pages 862--874. Springer,
  2007.

\bibitem{Bojanczyketal11}
M.~Boja\'nczyk, C.~David, A.~Muscholl, T.~Schwentick, and L.~Segoufin.
\newblock Two-variable logic on data words.
\newblock {\em ACM Trans. Comput. Log.}, 12(4), 2011.

\bibitem{Bojanczyketal09}
M.~Boja\'nczyk, A.~Muscholl, T.~Schwentick, and L.~Segoufin.
\newblock Two-variable logic on data trees and {XML} reasoning.
\newblock {\em J. ACM}, 56(3), 2009.

\bibitem{Bonnet11}
R.~Bonnet.
\newblock The reachability problem for vector addition system with one
  zero-test.
\newblock In {\em MFCS}, volume 6907 of {\em LNCS}, pages 145--157. Springer,
  2011.

\bibitem{deGroote&Guillaume&Salvati04}
P.~de~Groote, B.~Guillaume, and S.~Salvati.
\newblock Vector addition tree automata.
\newblock In {\em LICS}, pages 64--73, 2004.

\bibitem{Demrietal12}
S.~Demri, M.~Jurdzi\'nski, O.~Lachish, and R.~Lazi\'c.
\newblock The covering and boundedness problems for branching vector addition
  systems.
\newblock {\em J. Comput. Syst. Sci.}, 79(1):23--38, 2013.

\bibitem{Kosaraju82}
R.~Kosaraju.
\newblock Decidability of reachability in vector addition systems.
\newblock In {\em STOC}, pages 267--281, 1982.

\bibitem{Lambert92}
J.-L. Lambert.
\newblock A structure to decide reachability in {P}etri nets.
\newblock {\em Theor. Comput. Sci.}, 99(1):79--104, 1992.

\bibitem{Lazic10}
R.~Lazi\'c.
\newblock The reachability problem for branching vector addition systems
  requires doubly-exponential space.
\newblock {\em Inf. Process. Lett.}, 110(17):740--745, 2010.

\bibitem{Leroux12}
J.~Leroux.
\newblock Vector addition systems reachability problem (a simpler solution).
\newblock In {\em Turing-100}, volume~10 of {\em EPiC Series}, pages 214--228.
  EasyChair, 2012.

\bibitem{Lipton76}
R.~J. Lipton.
\newblock The reachability problem requires exponential space.
\newblock Technical Report~62, Dep. Comput. Sci., Yale Univ., Jan. 1976.

\bibitem{Mayr84}
E.~W. Mayr.
\newblock An algorithm for the general {P}etri net reachability problem.
\newblock {\em SIAM J. Comput.}, 13(3):441--460, 1984.

\bibitem{Rackoff78}
C.~Rackoff.
\newblock The covering and boundedness problems for vector addition systems.
\newblock {\em Theor. Comput. Sci.}, 6(2):223--231, 1978.

\bibitem{Rambow94}
O.~Rambow.
\newblock Multiset-valued linear index grammars: imposing dominance constraints
  on derivations.
\newblock In {\em ACL}, pages 263--270, 1994.

\bibitem{Reinhardt08}
K.~Reinhardt.
\newblock Reachability in {P}etri nets with inhibitor arcs.
\newblock In {\em RP}, volume 223 of {\em Electr. Notes Theor. Comput. Sci.},
  pages 239--264, 2008.

\bibitem{Schmitz10}
S.~Schmitz.
\newblock On the computational complexity of dominance links in grammatical
  formalisms.
\newblock In {\em ACL}, pages 514--524, 2010.

\bibitem{Stockmeyer74}
L.~J. Stockmeyer.
\newblock {\em The complexity of decision problems in automata theory and
  logic}.
\newblock PhD thesis, MIT, 1974.
\newblock {TR}-133, Lab. Comput. Sci.

\end{thebibliography}

\end{document}